\documentclass[aps,showpacs,twocolumn,prb,longbibliography,nobibnotes]{revtex4-1}

\usepackage[dvipdfmx]{graphicx}
\usepackage{dcolumn}
\usepackage{bm}
\usepackage{hyperref}
\usepackage[utf8]{inputenc}
\usepackage{pslatex}
\usepackage{textcomp} 
\usepackage{color}
\usepackage{lmodern} 
\usepackage[normalem]{ulem}
\usepackage{amsmath}
\usepackage{amsfonts}
\usepackage{graphicx} 
\graphicspath{{./figures/}}

\hypersetup{
    breaklinks = true,
    colorlinks = true,
    pdftitle = {}, 
    pdfauthor = {},
    pdfkeywords = {},
    linkcolor = blue,
    citecolor = blue,
    filecolor = black,
    urlcolor = magenta
}

\begin{document}

\title{Ab initio study of magnetocrystalline anisotropy, magnetostriction, and Fermi surface of L1$_0$ FeNi (tetrataenite) }

\author{Miros\l{}aw Werwi\'nski}\email[Corresponding author: ]{werwinski@ifmpan.poznan.pl}
\affiliation{Institute of Molecular Physics Polish Academy of Sciences, M. Smoluchowskiego 17, 60-179 Pozna\'{n}, Poland}

\author{Wojciech Marciniak}
\affiliation{Faculty of Technical Physics, Pozna\'{n} University of Technology, Pl. M. Skłodowskiej-Curie 5, 60-965 Pozna\'{n}, Poland}

\begin{abstract}
The ordered L1$_0$ FeNi phase (tetrataenite) is recently considered as a promising candidate for the rare-earth free permanent magnets applications.
In this work we calculate several characteristics of the L1$_0$ FeNi,
where most of the results come form the fully relativistic full potential FPLO method with the generalized gradient approximation (GGA).
A special attention deserves
the summary of the magnetocrystalline anisotropy energies (MAE's),
the full potential calculations of the anisotropy constant $K_3$, and
the combined analysis of the Fermi surface and three-dimensional $\mathbf{k}$-resolved MAE.
Other calculated parameters presented in this article are the 
magnetic moments $m_{s}$ and $m_{l}$,
magnetostrictive coefficient $\lambda_{001}$, 
bulk modulus B$_0$, and
lattice parameters.
%
%
The MAE's summary shows rather big discrepancies between the experimental MAE's from literature and also between the calculated MAE's.
The MAE's calculated in this work with the full potential and GGA are equal to 0.47~MJ~m$^{-3}$ from WIEN2k, 0.34~MJ~m$^{-3}$ from FPLO, and 0.23~MJ~m$^{-3}$ from FP-SPR-KKR code.
These last results strongly suggest that the value of MAE in GGA is below 0.5~MJ~m$^{-3}$.
It is also expected that this value is significantly underestimated due to the limitations of the GGA.
Unfortunately, as other authors suggest, even the MAE equal 1.3~MJ~m$^{-3}$ would be insufficient to raise the L1$_0$ FeNi from the category of semi-hard magnets.
However the L1$_0$ FeNi has still a potential to improve its MAE by modifications, like e.g. tetragonal strain or alloying.
The presented three-dimensional $\mathbf{k}$-resolved map of the MAE combined with the Fermi surface gives a complete picture of the MAE contributions in the Brillouin zone.
The calculated Fermi surface consists of closed hole pockets and open sheets.
It reflects a four-fold symmetry of the crystal and is closely related to the MAE($\mathbf{k}$).
The analysis of the effects of external factors, like strain, on the $\mathbf{k}$-resolved MAE and Fermi surface should be beneficial in engineering of the  hard magnetic properties.
The obtained from full potential FP-SPR-KKR method magnetocrystalline anisotropy constants $K_2$ and $K_3$ are several orders of magnitude smaller than the MAE/$K_1$ and 
equal to -2.0~kJ~m$^{-3}$ and 110~J~m$^{-3}$, respectively.
The calculated partial spin and orbital magnetic moments of the L1$_0$ FeNi are equal to 2.72 and 0.054~$\mu_{\mathrm{B}}$ for Fe and
0.53 and 0.039~$\mu_{\mathrm{B}}$ for Ni atoms, respectively.
The calculations of geometry optimization lead to a $c$/$a$ ratio equal to 1.0036, B$_0$ equal to 194~GPa, and $\lambda_{001}$ equal to 9.4~$\times$~10$^{-6}$.
\end{abstract}

\date{\today}

\maketitle

\section{Introduction}\label{sec:introduction}
%
%
The electric power generators, motors, and transformers are just a few examples where the magnetic materials find an application in the modern technology.
The hard magnetic materials used the most are the alnicos, hexaferrites, and Nd-Fe-B alloys.
The economical event from 2011 called \textit{The Rare-Earth Crisis}~\cite{bourzac_rare-earth_2011} destabilized i.a. the prices of neodymium 
motivating efforts to find new rare-earth free permanent magnets.
The ongoing search for hard magnetic materials free from rare-earth elements is summarized in several review articles.~\cite{kuzmin_towards_2014, mccallum_practical_2014, niarchos_toward_2015, hirosawa_current_2015, skomski_magnetic_2016, li_recent_2016}
Some of the promising candidates studied recently are e.g.
Fe/Co nanowires~\cite{bran_correlation_2015, palmero_enhanced_2016},
Fe-Co alloys doped with B and C~\cite{reichel_increased_2014, giannopoulos_structural_2015, salikhov_enhanced_2017,kuswik_perpendicularly_2017},
(Fe/Co)$_2$B~\cite{edstrom_magnetic_2015,dane_density_2015,belashchenko_origin_2015,wallisch_synthesis_2015},
(Fe/Co)$_5$XB$_2$~\cite{cedervall_magnetostructural_2016, werwinski_magnetic_2016, lamichhane_study_2016},
MnBi~\cite{antropov_magnetic_2014},
and the L1$_0$ phases such as FePt, CoNi, MnAl, and FeNi.~\cite{ayaz_khan_magnetocrystalline_2016, gutfleisch_fept_2005, edstrom_electronic_2014} 
The last candidate on the list is a subject of this work.
%
%
\begin{figure}[ht]
\includegraphics[width=0.55\columnwidth]{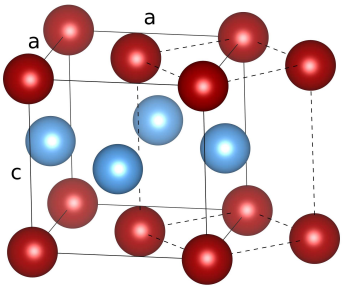}
\caption{\label{fig:struct} 
The L1$_0$ crystallographic structure. 
The solid lines designate a unit cell containing two formula units and the dashed lines confine a unit cell with a single formula.
}
\end{figure}
Existence of an ordered L1$_0$-type FeNi phase (tetrataenite) was confirmed in the sixties by N\'{e}el~\textit{et al.}~\cite{neel_magnetic_1964} in the study of a single crystal ordered by neutron bombardment and once again in the seventies by Paterson~\textit{et al.}~\cite{petersen_mossbauer_1977} in the study of taenite lamellae from the iron meteorite.
The L1$_0$ is the strukturbericht designation of the CuAu I-type ordered tetragonal phase.
The L1$_0$ unit cell confined by solid lines in Fig.~\ref{fig:struct} consists of two formula units.
Two faces of its  are occupied by one type of atoms.
The third face and corners are occupied by the second type of atoms.
A detailed study of crystallographic aspects of L1$_0$ magnetic materials can be found in a paper of Laughlin~\textit{et al.}~\cite{laughlin_crystallographic_2005}

%
In this work we investigate theoretically the magnetocrystalline anisotropy constants $K_2$ and $K_3$, Fermi surface, and bulk modulus of the tetrataenite.
This efforts are followed by the theoretical reinvestigation of the magnetostrictive coefficient $\lambda_{001}$ and magnetocrystalline anisotropy energy (MAE)
treated with a full potential fully relativistic method based on the full four component representation of the Bloch states.~\cite{eschrig_relativistic_2004}
Also the experimental~\cite{wu_spinorbit_1999, mizuguchi_artificial_2011, kotsugi_origin_2013, lewis_magnete_2014} and theoretical~\cite{wu_spinorbit_1999, miura_origin_2013, edstrom_electronic_2014} results from literature of the magnetocrystalline anisotropy of the L1$_0$ FeNi are considered.
One of the most important parameters from the perspective of permanent magnets application is a magnetocrystalline anisotropy constant $K_1$.
The measured values of $K_1$ of the L1$_0$ FeNi are relatively high~\cite{lewis_magnete_2014,poirier_intrinsic_2015} and equal up to 1.0--1.3~MJ~m$^{-3}$.
Skomski and Coey~\cite{skomski_magnetic_2016} suggest that even such high values of $K_1$ are insufficient to raise FeNi off the category of semi-hard magnets.
However it has been shown for the L1$_0$ FeNi films, that their intrinsic magnetic properties can be altered e.g. by engineering larger strains.~\cite{sakamaki_effect_2013,frisk_strain_2017}
Furthermore, the composition and microstructure of the L1$_0$ FeNi may be tailored as well to improve the FeNi potential for rare-earth-free permanent magnet application.~\cite{lewis_magnete_2014}
Skomski points out the beneficial self-organized microstructure of the L1$_0$ FeNi being reflected in a relatively high coercivity of about 120~mT.~\cite{zhukov_permanent_2016}
Some other characteristics indicating the L1$_0$ FeNi as a good candidate for hard magnets are magnetization approaching that of the Nd-Fe-B and relatively high Curie point near 550\textdegree~C~\cite{wasilewski_magnetic_1988}, however preceded by the critical temperature of the ordered state of about 320\textdegree~C.~\cite{neel_magnetic_1964}
Considering from the application point of view, a serious weakness of the FeNi remains achieving and retaining the L1$_0$ atomic order.~\cite{frisk_resonant_2016,frisk_strain_2017}

\section{Computational details}\label{sec:comp_details}
The determination from the first principles of the magnetocrystalline anisotropy constants and the magnetostrictive coefficient $\lambda_{001}$ requires the fully relativistic electronic band structure calculations.
The calculations are carried out by using the full-potential local-orbital minimum-basis scheme FPLO-14.0-49~\cite{eschrig_relativistic_2004,koepernik_full-potential_1999} with the generalized gradient approximation (GGA) in the Perdew-Burke-Ernzerhof (PBE) form.~\cite{perdew_generalized_1996} 
The calculations are performed up to a 80$^3$ \textbf{k}-mesh with tetrahedron method for integration, an energy convergence criterion $10^{-8}$~Ha, and a charge density convergence criterion $10^{-6}$.
%
%
The two-dimensional maps of MAE($\mathbf{k}$) are constructed on the 1000~\texttimes~1000 $\mathbf{k}$-mesh and the three-dimensional plot of MAE($\mathbf{k}$) is based on 250~\texttimes~250~\texttimes~170 $\mathbf{k}$-mesh within a selected one-eighth part of the full Brillouin zone.
An initial spin splitting is applied assuming the ferromagnetic structure. 
The bct representation of the L1$_0$ unit cell is used (as described e.g. by Edstr\"{o}m~\textit{et al.}~\cite{edstrom_electronic_2014}) with a space group $P$4/$mmm$ and atomic coordinates Fe~(0, 0, 0) and Ni~(0.5, 0.5, 0.5). 
Both the volume and $c$/$a$ ratio are optimized.
The calculated structural parameters are in good agreement with the corresponding experimental and theoretical values, see Table~\ref{tab:data}.
For the visualization of crystal structure the VESTA code~\cite{momma_vesta_2008} is used.
%
%
The magnetostriction is calculated with the same scheme as used by Wu~\textit{et al.}~\cite{wu_spinorbit_1999, wu_first_2001, zhang_mechanism_2011, wu_theory_2007}
In the latter references one can find the detailed description of the method but also the magnetostrictive coefficients calculated for several materials 
that stay in a good agreement with the experiment.
In order to calculate the magnetostrictive coefficient of the L1$_0$ FeNi we use the approach developed for cubic geometry as the tetragonal distortion in this system is very small (c/a = 1.0036).
To determine the magnetostrictive coefficient for a cubic material one has to calculate the strain dependences of the total energy ($E$) and the magnetocrystalline anisotropy energy (MAE), see Fig.~\ref{fig:dE_MAE}.
The dependence of the fractional change in length can be written based on the direction cosines of the magnetization ($\alpha$) 
and of the strain measurement ($\beta$):
\begin{equation}\label{eq:1}
\frac{\Delta l}{l_{0}}=\frac{3}{2}\lambda_{001}\left [ \sum_{i=1}^{3} \alpha^2_i\beta^2_i - \frac{1}{3} \right ] + 3\lambda_{111}\sum_{i\neq j}^{3}\alpha_i \alpha_j \beta_i  \beta_j.
\end{equation}
If the measurement is made along the [001] direction the equation simplifies to:
\begin{equation}\label{eq:2}
\frac{\Delta l}{l_{0}}=\frac{3}{2}\lambda_{001}\left [\alpha^2_z - \frac{1}{3} \right ]
\end{equation}
and for a single domain system it takes a form: 
\begin{equation}\label{eq:3}
\lambda_{001} = \frac{1}{3} \frac{l_{0}(\theta = 0^{\circ})-l_{0}(\theta = 90^{\circ}) }{l_{0}(\theta = 0^{\circ})+l_{0}(\theta = 90^{\circ})},
\end{equation}
where $\theta$ is the angle between the magnetization direction and the $c$ axis.
If the \textit{ab initio} calculated total energies are fitted in a quadratic form:
\begin{equation}\label{eq:4}
\begin{aligned}
E(\theta = 0^{\circ})= al^2+bl+c;\\
E(\theta = 90^{\circ})= al^2+bl+c+\mathrm{MAE},
\end{aligned}
\end{equation}
the equation for magnetostrictive coefficient can be written as:
\begin{equation}\label{eq:5}
\lambda_{001} = - \frac{2}{3} \frac{ \frac{\mathrm{d}(\mathrm{MAE})}{\mathrm{d}l}  }{b}.
\end{equation}
In this work the MAE's are evaluated based on the total energies calculated self-consistently for two perpendicular quantization axes:
\begin{equation}\label{eq:6}
\mathrm{MAE} = E(\theta = 90^{\circ}) -  E(\theta = 0^{\circ}).
\end{equation}

%
In addition to the FPLO calculations, the Korringa-Kohn-Rostoker (KKR) approach as implemented in the Munich SPR-KKR package (non-public full potential version 7.6.0) is used to calculate the magnetocrystalline anisotropy constant $K_3$.~\cite{ebert_et_al._munich_nodate, ebert_calculating_2011} 
The advantage of using full potential method for MAE calculations has been discussed before.~\cite{edstrom_magnetic_2015, ke_effects_2013}
The \textit{ab initio} calculations of $K_3$ are today still a numerically demanding task.
But what distinguishes the SPR-KKR among the other \textit{ab initio} codes is a numerical accuracy on the level of about 0.1~$\mu$eV in calculating total energy, which is the same order of magnitude as expected for the $K_3$ value of the L1$_0$ FeNi.
Another argument in favor of the SPR-KKR is that it has been successfully applied before to calculate $K_3$ for the magnetic shape memory Fe-Pd alloys.~\cite{kauffmann-weiss_enhancing_2012}
In order to get a converged values of $K_3$ for the L1$_0$ FeNi, the FP-SPR-KKR parameters 
of 10$^{-10}$~Ry energy convergence criterion and up to 225~\texttimes~225~\texttimes~158 $\mathbf{k}$-points (about 8 million) are necessary.
For Brillouin zone integration the special point method with a regular k-point grid is used.
Khan \textit{et al.} have shown for L1$_0$ FePt phase~\cite{ayaz_khan_magnetocrystalline_2016} that the KKR calculations of total energies are also quite sensitive to the angular momentum expansion $l_{max}$ cutoff used for the multipole expansion of the Green function.
Khan \textit{et al.} concluded that the angular momentum $l_{max}$~=~3 cuttoff yields to a qualitatively correct value of the MAE, however even for $l_{max}$~=~7 a full convergence is still difficult to reach.
Taking this conclusion into account we choose for our calculations a maximum angular momentum value $l_{max}$~=~4 (NL~=~5 in the KKR configuration file).
Our decision is further motivated by results of a convergence test of $K_1$ with respect to the angular momentum performed up to $l_{max}$~=~6. 
The test indicated that problems with convergence occur above the $l_{max}$~=~4 leading to divergence of the $K_1$ value.
This behavior may come from numerical problems in evaluating the Madelung potential and near-field corrections, pointed out by Khan~\textit{et~al.}~\cite{ayaz_khan_magnetocrystalline_2016}
The energy integrals are evaluated by contour integration on a circular energy path in complex plane (GRID~=~5), using 40 points of the E-mesh.
The calculations within the FP-SPR-KKR are carried out with the PBE exchange-correlation potential and with
the same crystallographic parameters of FeNi as used for FPLO MAE calculations.
The magnetocrystalline anisotropy energy in tetragonal crystal can be described by the following equation~\cite{buschow_physics_2004}:
\begin{equation}\label{eq:7}
\mathrm{MAE} = K_1 \sin^{2}\theta + K_2 \sin^4\theta + K_3 \sin^4\theta \cos 4\phi, 
\end{equation}
where K$_i$ are the anisotropy constants,
$\theta$ is the angle between the magnetization direction and the $c$ axis,
and $\phi$ is the angle between the magnetization and the $a$ axis within the basal plane of a tetragonal lattice.
For $\theta$~=~90$^{\circ}$ the Eq.~\ref{eq:7} takes the following form:
\begin{equation}\label{eq:8}
\mathrm{MAE} - K_1 - K_2 = K_3 \cos 4\phi. 
\end{equation}
The $K_3$ is then evaluated based on the total energies $E_{100}$ and $E_{110}$ calculated self-consistently for $\phi$ equal to 0$^{\circ}$ and 45$^{\circ}$ (directions [100] and [110] in the bct unit cell) from equation:
\begin{equation}\label{eq:10}
2 K_3 = E_{100} - E_{110}.
\end{equation}
The computational parameters used to obtain  K$_1$ and K$_2$ with FP-SPR-KKR are the same as presented above for the K$_3$ calculations, with an exception that lower number of  $\mathbf{k}$-points, about 1.5 million, is used in the whole Brillouin zone.

%
Furthermore the full-potential augmented plane-wave method FP-LAPW as implemented in the WIEN2k code~\cite{peter_blaha_wien2k_2014} is used to calculate a reference value of MAE.
Muffin-tin radii RMT are 2.29~a$_0$ for Fe and 2.29~a$_0$ for Ni atoms, where a$_0$ is Bohr radius.
The PBE exchange-correlation potential is used.
Plane wave cut-off parameter $RK_{max}$ is set to 10, which leads to above 245 basis functions.
Relativistic effects are included with the second variational treatment of spin-orbit coupling. 
The total energy convergence criterion is set to $10^{-8}$~Ry. 
The 55~\texttimes~55~\texttimes~39 \textbf{k}-points (about 120 thousand) are used in the whole Brillouin zone.

Although, the reproducibility of the results in density functional theory calculations of solids has been well established on the level of scalar relativistic estimation of the lattice parameters~\cite{lejaeghere_reproducibility_2016}, the accurate calculations of the magnetocrystalline anisotropy energy MAE and its derivatives, like magnetostrictive coefficient, remain a challenge.~\cite{ayaz_khan_magnetocrystalline_2016}
It is even harder to calculate the values of magnetocrystalline anisotropy constant $K_3$ which happens to be two orders of magnitude smaller than MAE.~\cite{kauffmann-weiss_enhancing_2012}
In this work we managed to calculate the $K_3$ of the L1$_0$ FeNi with the full potential thanks to the very accurate convergence tests.

\section{Density Functional Theory Calculations}\label{sec:dft}

The L1$_0$ FeNi phase has been studied \textit{ab inito} several times before.~\cite{wu_spinorbit_1999,ravindran_large_2001,miura_origin_2013,edstrom_electronic_2014}
In this work we reinvestigate the magnetocrystalline anisotropy energy MAE and the magnetostrictive coefficient $\lambda_{001}$ of the L1$_0$ FeNi phase and calculate its magnetocrystalline anisotropy constant $K_2$, $K_3$, Fermi surface, and bulk modulus.
Our results together with the literature data are summarized in Tab.~\ref{tab:data}.
\begin{table*}[!ht]
\begin{small}
\caption{\label{tab:data} The lattice parameters ($a$ and $c$), spin ($M_S$) and orbital ($M_L$) magnetic moments, magnetocrystalline anisotropy energies (MAE), magnetostrictive coefficients ($\lambda_{001}$), and bulk moduli (B$_0$) of the L1$_0$ FeNi phase.}
\centering
\begin{tabular}{l|ccccccccccc}
\hline \hline
quantity        				& $a$  	& $c$  	& M$_S$(Fe)		& M$_L$(Fe) 		& M$_S$(Ni) 		& M$_L$(Ni)		& MAE or $K_1$  		& MAE or $K_1$		&$\lambda_{001}$&B$_0$\\
unit        				& \AA{} & \AA{}	& $\mu_{\mathrm{B}}$	& $\mu_{\mathrm{B}}$ 	& $\mu_{\mathrm{B}}$ 	& $\mu_{\mathrm{B}}$ 	& $\mu$eV~formula$^{-1}$	& MJ m$^{-3}$	& 10$^{-6}$& GPa\\
\hline
experiment 				& 3.57~\cite{frisk_strain_2017} & 3.57~\cite{frisk_strain_2017} & 2.54\textpm0.16~\cite{cable_magnetic-moment_1973}	&  $\sim$0.05~\cite{kotsugi_origin_2013}		& 0.73\textpm0.04~\cite{cable_magnetic-moment_1973}	& 0.10~\cite{kotsugi_origin_2013} 	&  ---		& 0.58--1.3~\cite{mizuguchi_artificial_2011,lewis_magnete_2014}		& $\sim$9~\cite{bozorth_magnetic_1953}& ---\\	
FLAPW-GGA~\cite{wu_spinorbit_1999}	& 3.58	& 3.58	& 2.71		& 0.052			& 0.69			& 0.038			& 32		& 0.22		& 9.7	& ---\\
VASP-GGA~\cite{lewis_magnete_2014}	& ---	& ---	& ---		& ---			& ---			& ---			& 110		& 0.78		&---	& ---\\
VASP-GGA~\cite{miura_origin_2013}	& 3.556	& 3.584	& 2.65		& ---			& 0.61			& ---			& 78		& 0.56		&---	& ---\\
WIEN2k-GGA~\cite{miura_origin_2013}	& ---	& ---	& ---		& ---			& ---			& ---			& 69		& 0.48		&---	& ---\\
WIEN2k-GGA~\cite{edstrom_electronic_2014}& 3.56	& 3.58 	& 2.69		& ---			& 0.67			& ---			& 69		& 0.48		&---	& ---\\
WIEN2k-GGA (this work)			& ---	& --- 	& 2.69		& 0.052			& 0.66			& 0.036			& 67		& 0.47		&---	& ---\\
ASA-SPR-KKR-GGA~\cite{edstrom_electronic_2014}& --- & ---& 2.73		& ---			& 0.62			& ---			& 110		& 0.77		&---	& ---\\	
FP-SPR-KKR-GGA (this work) 		& --- 	& ---	& 2.69		& 0.053			& 0.60			& 0.036			& 32  		& 0.23 		& --- & ---\\
FPLO14-GGA (this work) 			& 3.56 	& 3.58	& 2.72		& 0.054			& 0.53			& 0.039			& 48  		& 0.34 		& 9.4 & 194\\
\hline \hline
\end{tabular}              
\end{small}
\end{table*}

\subsection{Bulk Modulus}

%
\begin{figure}[ht]
\includegraphics[trim = 310 10 10 25,clip,height=\columnwidth,angle=270]{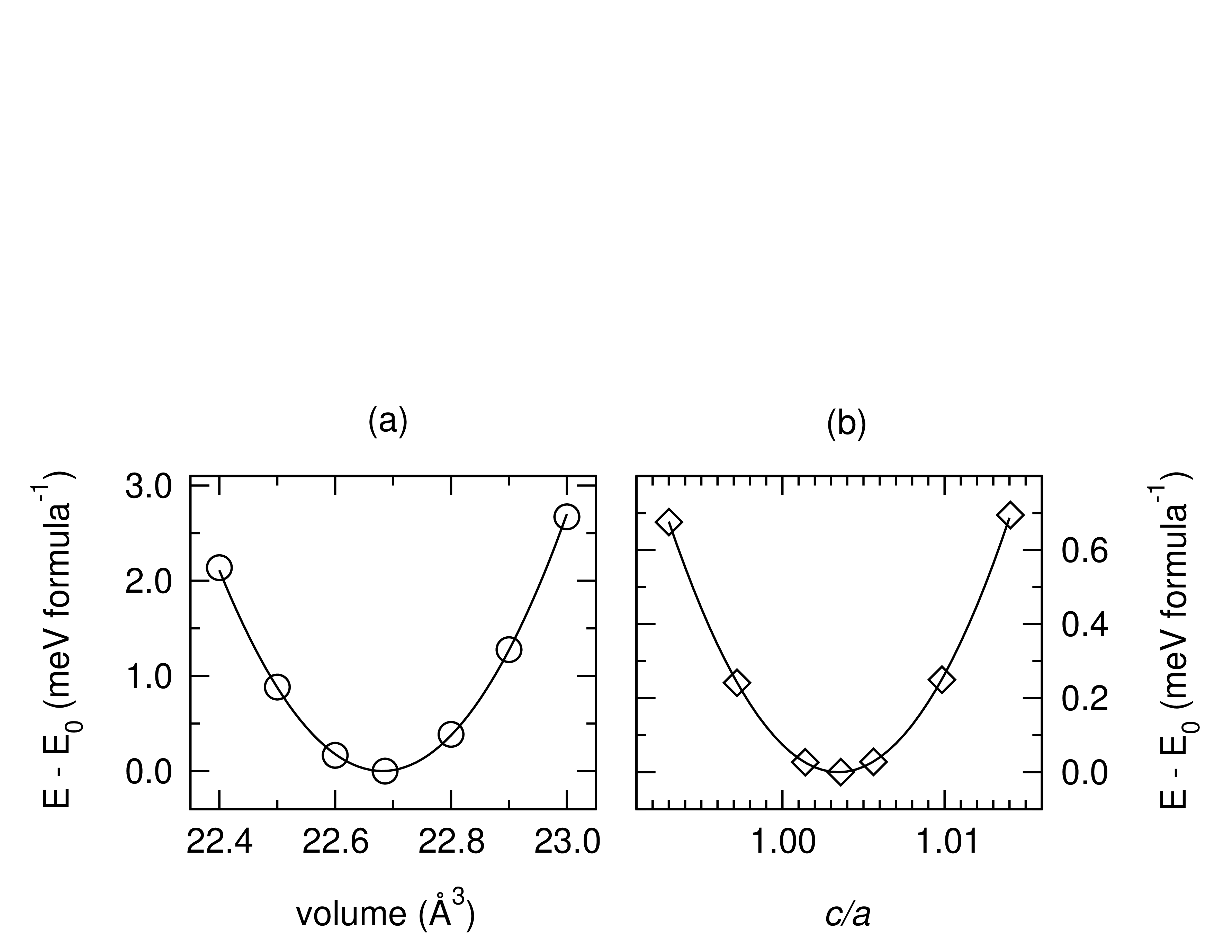}
\caption{\label{fig:vol_coa} 
(a) The volume dependence of the total energy and (b) total energy \textit{versus} the lattice parameters $c$/$a$ ratio as calculated for the L1$_0$ FeNi with the FPLO14 PBE+so method. 
The $V$($E$) is calculated with a fixed $c$/$a$~$\sim$~1.0056 as obtained previously by Edst\"{o}m~\textit{et al.}~\cite{edstrom_electronic_2014}
The equilibrium volume equals 22.686~\AA{}$^3$. 
The equilibrium $c$/$a$ ratio calculated with a fixed volume 22.686~\AA{}$^3$ equals 1.0036.
The energy scale is shifted so that the energy minimum equals zero.
}
\end{figure}
We start from the calculations of the equilibrium volume (22.686~\AA{}$^3$), followed by the calculations of equilibrium $c$/$a$ ratio (1.0036), see Fig.~\ref{fig:vol_coa}.
The optimization leads to a tetragonal structure with the lattice parameters $a=3.56$~\AA{} and $c=3.58$~\AA{}.
The energy-volume data, see Fig.~\ref{fig:vol_coa}~(a), allows to calculate the bulk modulus B$_0$ for the L1$_0$ FeNi by fitting the third-order Birch-Murnaghan equation of state.~\cite{birch_finite_1947}
It leads to a B$_0$ equal to 194~GPa at 0~K.
The previous calculations with coherent potential approximation CPA, of the Fe$_{0.5}$Ni$_{0.5}$ random alloy have given the B$_0$ value of about 220~GPa.~\cite{abrikosov_theoretical_1995}
For comparison, the experimental values of B$_0$ for the Fe$_{0.5}$Ni$_{0.5}$ alloy measured at room temperature vary between 165~GPa and 177~GPa~\cite{wijn_1.2.1.2.9_1986}
and the experimental B$_0$ values for Fe and Ni are about 170~GPa and 180~GPa, respectively.

\subsection{Magnetostrictive Coefficient $\lambda_{001}$}

%
\begin{figure}[ht]
\includegraphics[trim = 80 10 10 20,clip,height=0.9\columnwidth,angle=270]{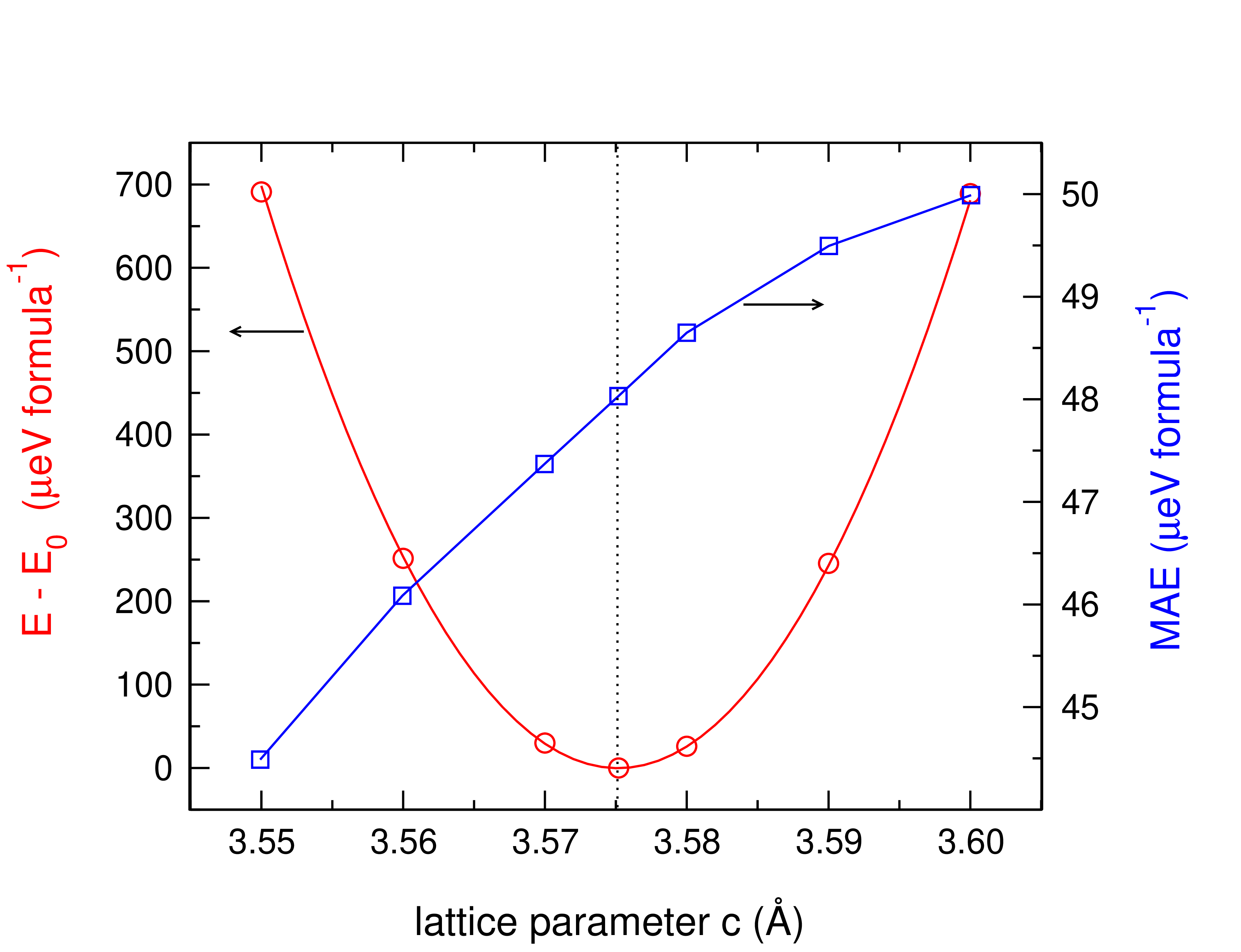}
\caption{\label{fig:dE_MAE} 
The calculated total energy and magnetocrystalline anisotropy energy of the L1$_0$ FeNi \textit{versus} the length of the lattice parameter $c$. 
The results are obtained within the FPLO14 PBE+so method.
A constant volume mode is adopted for the distortion. 
The dotted line indicates the equilibrium lattice parameter.
}
\end{figure}

The total and magnetocrystalline anisotropy energies \textit{versus} the length of the lattice parameter $c$ are presented in Fig.~\ref{fig:dE_MAE} as necessary to evaluate the magnetostrictive coefficient $\lambda_{001}$ from Eq.~\ref{eq:5}.
The total energy and MAE are calculated based on the optimized crystal structure in a constant volume mode adopted for the distortion.
The MAE calculated for the equilibrium L1$_0$ FeNi structure equals 48~$\mu$eV~formula$^{-1}$ (0.34~MJ~m$^{-3}$).
Initial tests performed with volume relaxation for 50$^3$~\textbf{k}-points have shown that the constant volume mode underestimates the $\lambda_{001}$ of the L1$_0$ FeNi by about 10\%.
For the price of this inaccuracy, in constant volume mode we can perform calculations up to 80$^3$~\textbf{k}-points.
The resultant $\lambda_{001}$ obtained for 60$^3$, 70$^3$, and 80$^3$~\textbf{k}-points are equal to 11.3~$\times$~10$^{-6}$, 8.7~$\times$~10$^{-6}$, and 9.4~$\times$~10$^{-6}$, respectively.
The variation of the estimated $\lambda_{001}$ with number of \textbf{k}-points comes from the numerical inaccuracy in the evaluation of a very small quantity as MAE.
Nevertheless, the calculated here $\lambda_{001}$~=~9.4~$\times$~10$^{-6}$ stays in a good agreement with the previous theoretical result by Wu and Freeman~\cite{wu_spinorbit_1999}, $\lambda_{001}$~=~9.7~$\times$~10$^{-6}$, and with the experimental value by Bozorth~\cite{bozorth_magnetic_1953} $\lambda_{100} \sim 9 \times 10 ^{-6}$.
$\lambda_{001} \sim 10 \times 10^{-6}$ for the L1$_0$ FeNi is rather small value.
It is of the same order of magnitude as for elements Fe and Ni and three orders of magnitude smaller than for the magnetostrictive material Terfenol-D.

\subsection{Magnetic Moments}
%
The spin magnetic moments calculated with the FPLO for the L1$_0$ FeNi are equal 2.72~$\mu_{\mathrm{B}}$ for the Fe atom and 0.53~$\mu_{\mathrm{B}}$ for the Ni atom, see Tab.~\ref{tab:data}. 
The calculated spin magnetic moments on Fe and Ni stay in relatively good agreement with the values measured on Fe equal to 2.54\textpm0.16~$\mu_{\mathrm{B}}$ and on Ni equal to 0.73\textpm0.04~$\mu_{\mathrm{B}}$ for the L1$_0$ FeNi phase.~\cite{cable_magnetic-moment_1973}
The calculated orbital magnetic moments are equal to 0.054~$\mu_{\mathrm{B}}$ for the Fe atom and 0.039~$\mu_{\mathrm{B}}$ for the Ni atom. 
They are also close to the experimental values on Fe equal to $\sim$0.05~$\mu_{\mathrm{B}}$ and on Ni equal to 0.10~$\mu_{\mathrm{B}}$ as measured for the L1$_0$ FeNi phase.~\cite{kotsugi_origin_2013}
The calculated and measured orbital magnetic moments on Fe in the L1$_0$ FeNi are reduced in comparison to the experimental value of 0.086~$\mu_{\mathrm{B}}$ for the bcc iron.~\cite{chen_experimental_1995}
The calculated orbital magnetic moments on Ni (0.039~$\mu_{\mathrm{B}}$) in the L1$_0$ FeNi are also reduced in respect to the experimental value of 0.055~$\mu_{\mathrm{B}}$ for the fcc nickel.~\cite{mook_magnetic_1966}
The calculated total magnetic moment is equal to 3.34~$\mu_{\mathrm{B}}$~formula$^{-1}$ (1.67~$\mu_{\mathrm{B}}$~atom$^{-1}$) which is not particularly high value for 3$d$-based magnetic materials.
From the perspective of hard magnetic materials this reduction of magnetic moment in comparison to e.g. pure bcc Fe is beneficial for the magnetic hardness but adversely affects the energy product.

\subsection{Magnetocrystalline Anisotropy Energy}

Getting consistent MAE results from different first principles codes is still a challenge.
The difficulties come from such factors as a complex shape of the valence band structure or a demand of very high numerical accuracy.~\cite{ayaz_khan_magnetocrystalline_2016}
The differences between the results from various codes may come from application of different approximations like the atomic sphere approximation, the lack of crystal structure optimization, or the insufficient number of \textbf{k}-points.~\cite{edstrom_magnetic_2015,kuzmin_magnetic_2015,ayaz_khan_magnetocrystalline_2016}
Some recent papers discuss however the reproducibility of the MAE between WIEN2k and KKR methods~\cite{ayaz_khan_magnetocrystalline_2016} and between FPLO and WIEN2k.~\cite{edstrom_magnetic_2015}

The calculated in GGA MAE's of the L1$_0$ FeNi taken from literature are equal to 0.22~\cite{wu_spinorbit_1999}, 0.48~\cite{edstrom_electronic_2014}, 0.56~\cite{miura_origin_2013}, and 0.78~\cite{lewis_magnete_2014}~MJ~m$^{-3}$.
The experimental determination of the L1$_0$ FeNi magnetocrystalline anisotropy constant $K_1$ is ambiguous as well, with a spread in the $K_1$ values from 0.58~\cite{mizuguchi_artificial_2011}, trough 0.67~\cite{neel_magnetic_1964}, up to 1.3~\cite{lewis_magnete_2014}~MJ~m$^{-3}$.
The MAE's of the L1$_0$ FeNi calculated in GGA in this work are equal to 0.23, 0.34, and 0.47~MJ~m$^{-3}$ from FP-SPR-KKR, FPLO, and WIEN2k, respectively.

Based on the above theoretical results we expect that the accurate GGA value of the MAE for the L1$_0$ FeNi is below 0.5~MJ~m$^{-3}$.
Such value could be an argument for removing the L1$_0$ FeNi from a list of candidates for rare-earth free permanent magnets.
Some authors suggest however that bare GGA is insufficient for describing the L1$_0$ FeNi and a consideration of orbital polarization corrections is necessary.~\cite{ravindran_large_2001,miura_origin_2013}
It has been shown that inclusion of the orbital polarization causes a significant increase of the MAE of the L1$_0$ FeNi (from $\sim$0.55 to $\sim$1.23~MJ~m$^{-3}$)~\cite{ravindran_large_2001} and (from 0.48 to 0.84~MJ~m$^{-3}$)~\cite{miura_origin_2013}.
Unfortunately, even the MAE of about 1~MJ~m$^{-3}$ may be insufficient to raise the L1$_0$ FeNi from the category of semi-hard magnets.~\cite{skomski_magnetic_2016}
To reach that goal further efforts on increasing the magnetocrystalline anisotropy have to be made.

\subsection{Magnetocrystalline Anisotropy Constant $K_3$}
%
The magnetocrystalline anisotropy constant $K_3$ of the L1$_0$ FeNi is particularly interesting from a perspective of spintronic applications of the material, e.g. for the magnetic tunnel junction.~\cite{miura_origin_2013}
In a tetragonal crystal the $K_3$ can be defined by Eq.~\ref{eq:7}. 
Unfortunately we cannot calculate the $K_3$ with the FPLO code, which was applied to obtain previous results.
To get the $K_3$ we use then the FP-SPR-KKR package which produces the value of $K_3$ equal to 110~J~m$^{-3}$.
It is four orders of magnitude smaller from the value of MAE or $K_1$ for the L1$_0$ FeNi. 
For comparison, the values of $K_3$ calculated in ASA-SPR-KKR for magnetic shape memory alloys Fe-Pd~\cite{kauffmann-weiss_enhancing_2012} have similar order of magnitude ($\sim10^3$~J~m$^{-3}$) as the one presented above.
The details of calculation method and the motivations for using FP-SPR-KKR package for evaluating $K_3$ are presented in the introductory section.
\begin{figure}[ht]
\includegraphics[trim = 330 80 10 30,clip,height=0.98\columnwidth,angle=270]{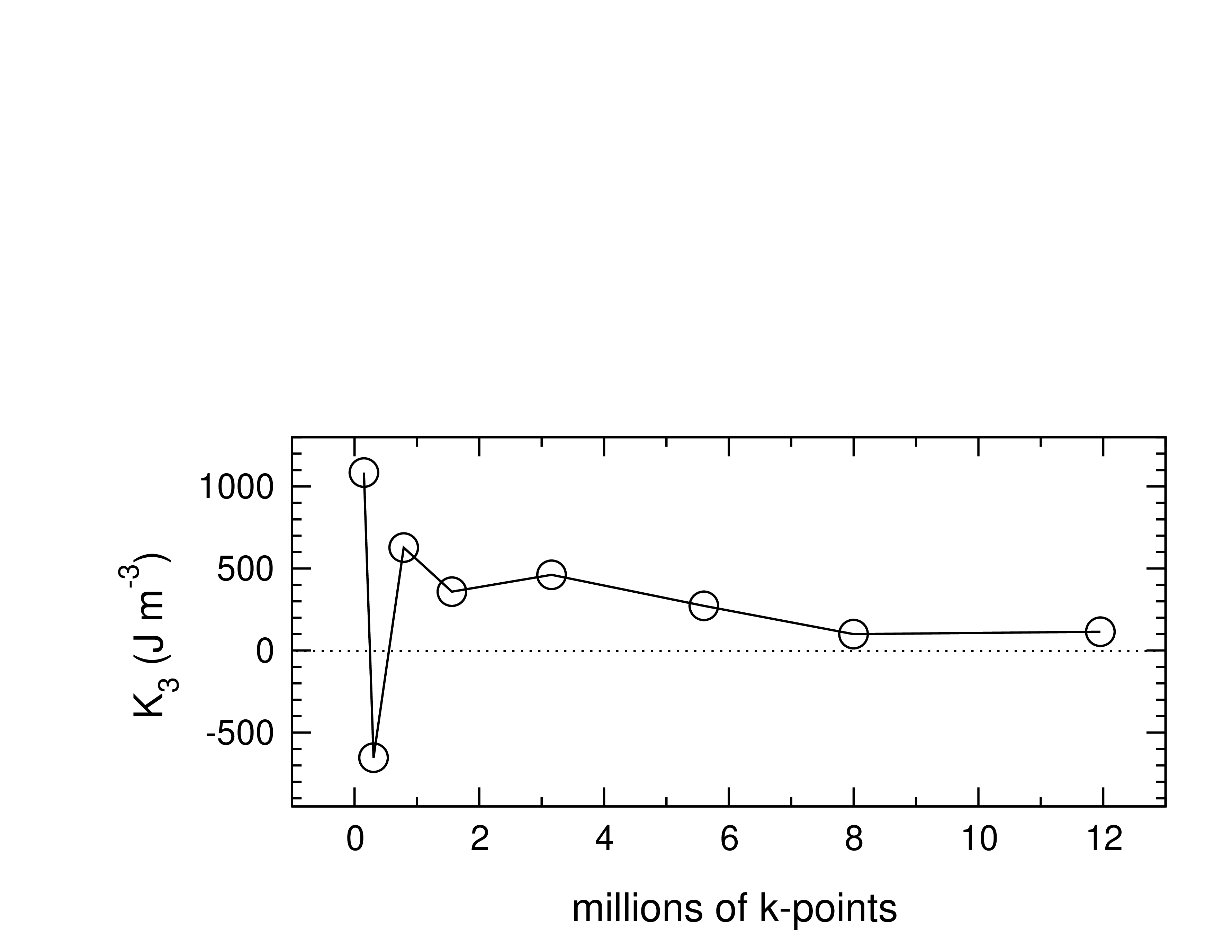}
\caption{\label{fig:k3_vs_k} 
The magnetocrystalline anisotropy constant $K_3$ of the L1$_0$ FeNi phase \textit{versus} the number of $\mathbf{k}$-points. 
The full potential FP-SPR-KKR code with the PBE approximation is used.
}
\end{figure}
As the $K_3$ is $\mathbf{k}$-mesh sensitive the convergence test is made, see Fig.~\ref{fig:k3_vs_k}.
It can be noticed that large number of $\mathbf{k}$-points is necessary to get a satisfactory convergence of $K_3$.
\begin{figure}[ht]
\includegraphics[trim = 310 20 20 100,clip,height=0.98\columnwidth,angle=270]{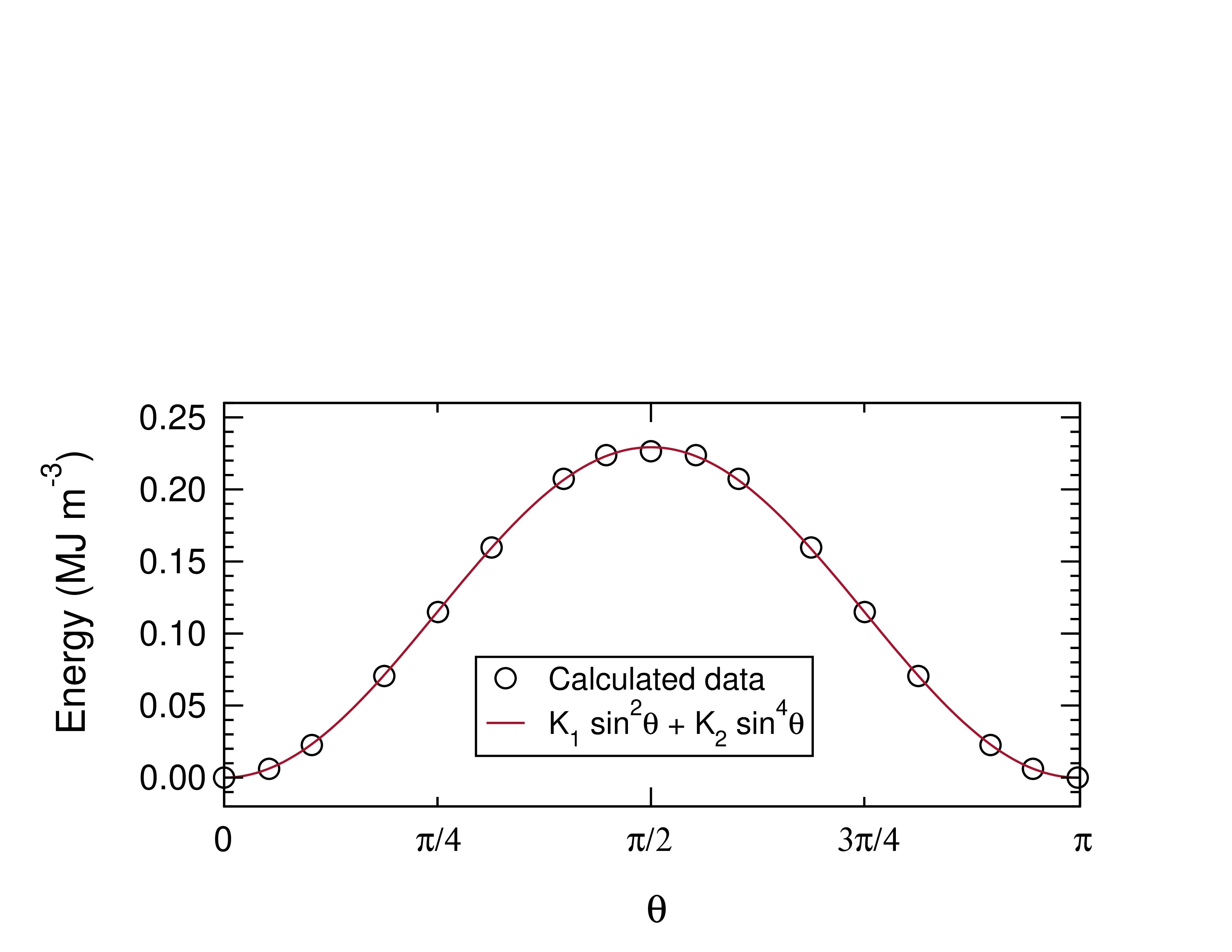}
\caption{\label{fig:k2} 
Energy as a function of the polar angle $\theta$ between the magnetization direction and the $c$ axis for the L1$_0$ FeNi. $K_1$~=~0.23~MJ~m$^{-3}$, $K_2$~=~-2.0~kJ~m$^{-3}$.
The full potential FP-SPR-KKR code with the PBE approximation is used.
}
\end{figure}
To complete the FP-SPR-KKR analysis also the K$_1$ and K$_2$ magnetocrystalline anisotropy constants are calculated.
In Fig.~\ref{fig:k2} we present the energy dependence as a function of the polar angle $\theta$ between the magnetization direction and the $c$ axis.
Parameters from a fit to $K_1 \sin^{2}\theta + K_2 \sin^4\theta$ are $K_1$~=~0.23~MJ~m$^{-3}$ and $K_2$~=~-2.0~kJ~m$^{-3}$, see Eq.~\ref{eq:8}.
The lowest energy in Fig.~\ref{fig:k2} corresponds to [001] quantization axis and the highest energy to [100] axis.

\subsection{Fermi Surface}

%
\begin{figure}[ht]
\includegraphics[width=0.9\columnwidth]{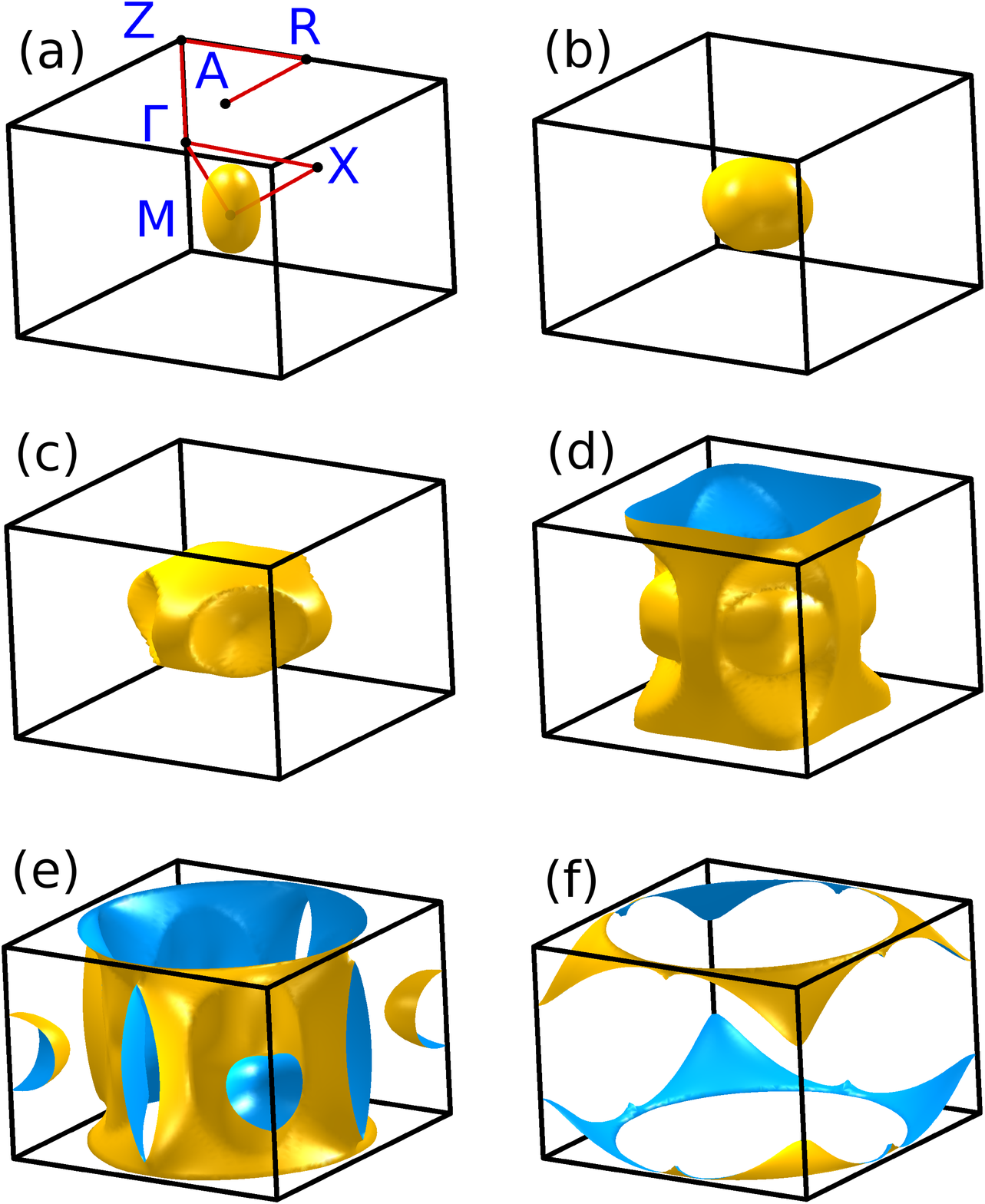}
\caption{\label{fig:fs} The Fermi surface of the L1$_0$ FeNi obtained within the FPLO14 PBE+so method.}
\end{figure}
%
%
Figure~\ref{fig:fs} presents the Fermi surface of the L1$_0$ FeNi.
The tetragonal crystal structure of the L1$_0$ FeNi is reflected in a uniaxial anisotropy and a four-fold symmetry of Fermi surface sheets.
The seven sheets are divided into two groups of closed hole pockets and open sheets.
The nested hole pockets (a), (b), and (c) are centered at the high symmetry point M.
One another hole pocket is centered at point $\Gamma$, see panel (e).
When the hole pockets permit for only closed orbits 
the mainly open sheets (d), (e), and (f) allow for both the opened and closed orbits.

%
We have calculated the Fermi surface of the L1$_0$ FeNi both as the basic characteristic of the material and 
as the basis for a {\bf k}-resolved MAE analysis which will be presented in the next paragraph.
Basic knowledge on the Fermi surface of the crystal lets us think on manipulating of its properties using the emerging method called the Fermi surface engineering.
The method makes relations between a shape of the Fermi surface and external factors like doping or strain.
The Fermi surface engineering technique has been successfully applied e.g. to 
transparent conductors~\cite{katayama-yoshida_engineering_2003} 
and 
superconductors.~\cite{hsu_manipulating_2016}

\subsection{MAE Analysis in Reciprocal Space}
%

%
Most often the anisotropic magnetic properties of the materials are analyzed in the real space.
The spin and orbital moments and magnetocrystalline anisotropy constants are presented as functions of the atomic position or  angle.~\cite{miura_origin_2013,antropov_magnetic_2014,thersleff_towards_2017}
The magnetocrystalline anisotropy energy (MAE) is one of the most important intrinsic properties of the hard magnetic crystals which can be analyzed in reciprocal space as a $\mathbf{k}$-resolved quantity.
%
%
The MAE can be determined with the magnetic force theorem~\cite{liechtenstein_local_1987,wang_validity_1996,wu_spinorbit_1999} from a formula:
\begin{multline}\label{eq:10}
\mathrm{MAE} = E(\theta = 90^{\circ}) -  E(\theta = 0^{\circ}) =\\
= \sum_{occ'} \epsilon_{i}(\theta = 90^{\circ}) -  \sum_{occ''} \epsilon_{i}(\theta = 0^{\circ}) + O(\delta \rho^n), 
\end{multline}
where $\epsilon_{i}$ is the band energy of the $i$th state and $\theta$ is the angle between the magnetization direction and the $c$ axis.
The modern computer technique allows to visualize the MAE($\mathbf{k}$) structure in a three-dimensional (3D) Brillouin zone.
The 3D maps of magnetocrystalline anisotropy have been recently calculated for (Fe/Co)$_2$B alloys~\cite{belashchenko_origin_2015} -- another candidates for the rare-earth free permanent magnets.
However usually the MAE($\mathbf{k}$) dependences are presented along one-dimensional $\mathbf{k}$-path.~\cite{wu_spinorbit_1999,wu_theory_2007,edstrom_magnetic_2015}
%
%
Although various authors attempt to interpret the MAE($\mathbf{k}$) without a 3D-resolution,
in our opinion the $\mathbf{k}$-path or single $\mathbf{k}$-point analyses do not cover properly the complexity of magnetocrystalline anisotropy.
We think that in order to get a full picture of it the entire Brillouin zone should be considered.~\cite{edstrom_magnetic_2015}
%
%
For a hard magnetic material 
the calculated 3D-MAE($\mathbf{k}$) plot is a unique characteristic 
as for example the Fermi surface is for a metal.
Interesting is that the connections between the 3D-MAE($\mathbf{k}$) and the Fermi surface go beyond the above comparison
and a fact is a close relation between these two.
Both, the 3D-MAE($\mathbf{k}$) and Fermi surface, are reciprocal space electronic structure characteristic. 
Furthermore, as the Fermi level is an upper integration boundary of the MAE, the sheets of the Fermi surface indicate the borders between the distinct regions of the MAE.
Because of that the  Fermi surface is a link between the calculated 3D-MAE($\mathbf{k}$) and experiment.
The 3D-MAE($\mathbf{k}$) analysis is relevant for hard magnetic materials also because it can determine directions to improve the MAE. 
In relation to the Fermi surface engineering an improvement technique based on the MAE($\mathbf{k}$) analysis could be called magnetocrystalline anisotropy engineering. 
%

%
\begin{figure}[ht]
\includegraphics[width=0.99\columnwidth]{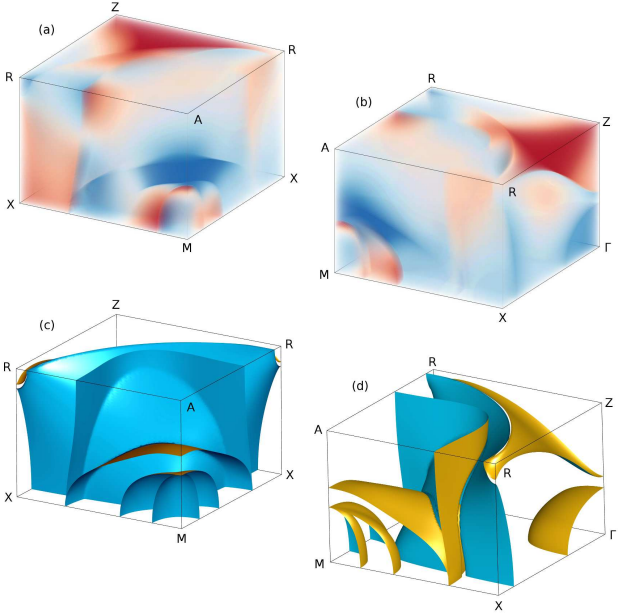}
\caption{\label{fig:MAE_3d_vs_fs}
(c), (d): The Fermi surface of the L1$_0$ FeNi view from two perspectives, 
together with corresponding (a), (b): {\bf k}-resolved contributions to MAE. 
The MAE({\bf k}) results are obtained by the magnetic force theorem within the FPLO14 PBE+so method. 
Only a reducible one-eighth part of the full Brillouin zone is presented. 
The magnitude of negative and positive values is indicated by the intensity of blue and red color, respectively.
}
\end{figure}
%
%
In Fig.~\ref{fig:MAE_3d_vs_fs} we present the calculated 3D-MAE($\mathbf{k}$) plots of the L1$_0$ FeNi.
We can see that the whole Brillouin zone is filled by the positive (red) and negative (blue) contributions. 
The magnitude of the $\mathbf{k}$-resolved contributions is in order of $10^{-3}$~eV per $\mathbf{k}$-point,
where the value of the total MAE is three orders of magnitude smaller (48~$\mu$eV~formula$^{-1}$).
A comparison between the MAE contributions and total value indicates a fine compensation of the relatively large negative and positive components.
From Fig.~\ref{fig:MAE_3d_vs_fs} it is easy to notice that the overall shape of the MAE($\mathbf{k}$) is dictated by the Fermi surface.
The Fermi surface sheets divide the Brillouin zone into several mainly positive or negative regions.
The two most prominent positive contributions come from (1) the vicinity of the R-Z line along the [100] quantization axis and (2) from the spherical region around the M point.
Actually, around the M point we observe a sequence of negative, positive, and again negative regions filling the volumes between the corresponding nested hole pockets of the Fermi surface. 
A similar alternating ordering of the $\mathbf{k}$-resolved contributions to anisotropy constant $K$ has been presented for the (Fe/Co)$_2$B alloys.~\cite{belashchenko_origin_2015}

%
\begin{figure}[ht]
\includegraphics[width=0.99\columnwidth]{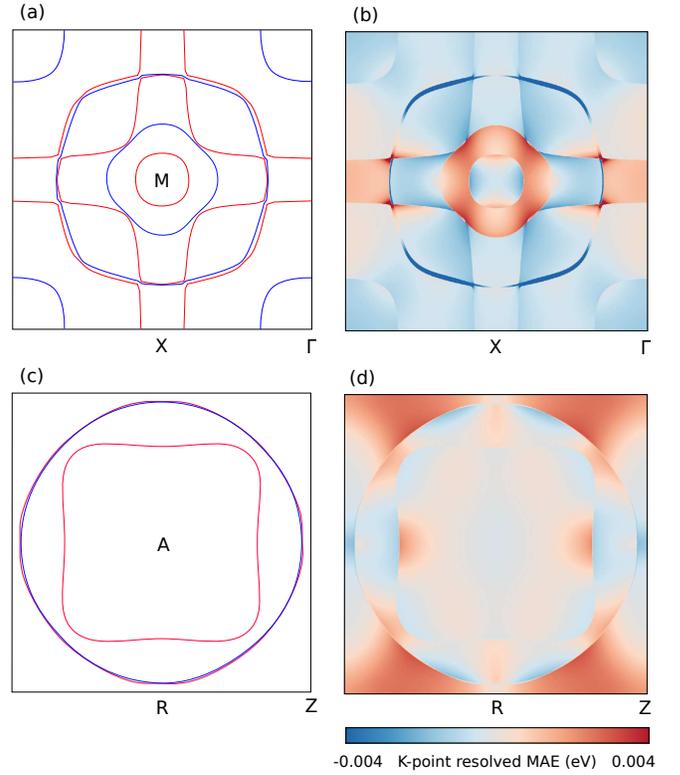}
\caption{\label{fig:slabs} (a), (c): The $\Gamma$-X-M and R-Z-A cross-sections of the Fermi surface of the L1$_0$ FeNi, together with the corresponding (b), (d): cross-sections of MAE contributions of each {\bf k}-point. The results are obtained by the magnetic force theorem within the FPLO14 PBE+so method. The different colors of Fermi surface contours are used as the guide for the eye only.
}
\end{figure}
In Fig.~\ref{fig:slabs} the cross-sections of the MAE({\bf k}) for planes $\Gamma$-X-M and R-Z-A are shown together with corresponding cross-sections of the Fermi surface.
At these plots it is even easier to notice how the Fermi surface splits the regions with differing MAE({\bf k}) components.
The MAE cross-sections also confirm that the first hole pocket around the M point contains mainly negative $\mathbf{k}$--resolved contributions and the regions between the next two hole pockets are respectively positive and negative. 
Although the MAE($\mathbf{k}$) structure reflects the uniaxial anisotropy of the crystal, 
it is easy to notice how the four-fold symmetry is broken.
A reason for this is that the [100] direction has been distinguished as a quantization axis.

%
The above-described MAE($\mathbf{k}$) analysis reveals details of the structure of magnetocrystalline anisotropy of the L1$_0$ FeNi
which are the distribution of positive and negative components, the magnitude of these shares, or the detailed relationship between the MAE($\mathbf{k}$) and the Fermi surface.
Unfortunately, an extra fine compensation of the large MAE($\mathbf{k}$) contributions makes it difficult to predict the roads to increase the MAE of the L1$_0$ FeNi.
Because of the compensation effect, even large changes of the MAE($\mathbf{k}$) structure will finish as hard to predict small changes to the MAE.
Based on the combined 3D-MAE and Fermi surface analysis we are able however to point out the sheets of Fermi surface which should be extended or reduced to enlarge the regions of positive MAE contributions.
Unfortunately, this simple line of reasoning is difficult to realize in practice.
Taking into account the innovative nature of the combined 3D-MAE and Fermi surface analysis, we do not exclude that this approach will be more fruitful for another hard magnetic material.

\section{Summary and conclusions}
%

%
The magnetocrystalline anisotropy energy MAE of the L1$_0$ FeNi calculated in this work within GGA goes below 0.5~MJ~m$^{-3}$.
It has been shown in literature that the more reliable model including orbital polarization corrections gives about twice this value,
whereas the experimental values of the anisotropy constant $K_1$ from literature oscillate around 1.0~MJ~m$^{-3}$.
Regarding application of the L1$_0$ FeNi as rare-earth free permanent magnets,
the above numbers of MAE/$K_1$ are rather discouraging.
Other known limitations of tetrataenite are 
practical difficulties with synthesis of the ordered phase and 
relatively low critical temperature of the ordered state.   
However the L1$_0$ FeNi has still a potential to improve its MAE by modifications, like e.g. a tetragonal strain or an interstitial or substitutional alloying.
In favor of the L1$_0$ FeNi speak also the high saturation magnetization and Curie temperature.
The L1$_0$ FeNi with well defined anisotropic parameters may also find applications in modern electronics.
From point of view of computations 
it can be also used as a reference standard for advanced \textit{ab initio} calculations 
due to relatively simple crystal structure.

%
In the reciprocal space analysis of the hard magnetic materials it is common to present the MAE($\mathbf{k}$) contributions together with the bandstructure along the $\mathbf{k}$-path.
In this work we have shown the three-dimensional $\mathbf{k}$-resolved analysis of the MAE (3D-MAE) which gives a complete picture of the MAE contributions in the Brillouin zone.
Unfortunately, in case of the L1$_0$ FeNi we cannot use the $\mathbf{k}$-resolved analysis to give specific suggestions.
In return we show the close interconnection between the 3D-MAE and the Fermi surface.
We expect that analysis of the effect of strain or alloying on the $\mathbf{k}$-resolved MAE may allow to improve the properties of hard magnetic materials.

%
The calculated 
spin and orbital magnetic moments
and magnetostrictive coefficient $\lambda_{001}$ 
stay in an acceptable agreement with the previous theoretical results and measurements.
The calculated bulk modulus $B_0$, magnetocrystalline anisotropy constants $K_2$ and $K_3$, and Fermi surface require experimental confirmation.

%
The conducted full relativistic calculations can be also used to formulate several general conclusions regarding the computational parameters.
The fact that that spin-orbit coupling is the origin of the orbital magnetic moment, magnetocrystalline anisotropy, and magnetostriction is well known.
In order to accurately calculate the magnetic parameters it is also crucial to consider the full potential.
The significant discrepancies between the calculated MAE's even from various full potential GGA implementations show how sensitive the calculations are.
To improve the reliability of the results we suggest to choose the computational parameters very carefully and use a second code for cross-checking.
Though it is well known that the orbital magnetic moment on Fe is underestimated in GGA, 
it is not obvious that it leads to underestimation of the MAE.
For better description of the magnetic moments in considered system the orbital polarization corrections and dynamical mean field theory can be used.
In this work we have also shown that it is technically possible to calculate the anisotropy constant $K_3$ with full potential.
The calculated value of $K_3$ is four orders of magnitude smaller than the MAE 
and a huge number of $\mathbf{k}$-points is indispensable to get a consistent value.

\section{Acknowledgements}
We thank A. Edstr\"{o}m and B. Wasilewski for discussions. 
We also acknowledge the financial support from the Foundation of Polish Science grant HOMING. 
The HOMING programme is co-financed by the European Union under the European Regional Development Fund.

\bibliography{feni_paper}       

\end{document}